\documentclass[graybox]{svmult}

\usepackage{mathptmx}       % selects Times Roman as basic font
\usepackage{helvet}         % selects Helvetica as sans-serif font
\usepackage{courier}        % selects Courier as typewriter font
\usepackage{type1cm}        % activate if the above 3 fonts are
\usepackage{makeidx}         % allows index generation
\usepackage{graphicx}        % standard LaTeX graphics tool
\usepackage{multicol}        % used for the two-column index
\usepackage[bottom]{footmisc}% places footnotes at page bottom

\makeindex             % used for the subject index
\begin{document}

\title*{Measuring baryon acoustic oscillations with angular two-point correlation function}
\author{Jailson S. Alcaniz$^{*}$, Gabriela C. Carvalho, Armando Bernui, Joel C. Carvalho \& Micol Benetti}
\authorrunning{J. S. Alcaniz et al.} 
\institute{$^{*}$\email alcaniz@on.br \at 
Coordena\c{c}\~ao de Astronomia \& Astrof\'{\i}sica, Observat\'orio Nacional, 20921-400, Rio de Janeiro -- RJ, Brazil}

\maketitle

\abstract{The Baryon Acoustic Oscillations (BAO) imprinted a characteristic correlation length in the large-scale structure of the universe that can be used as a standard ruler for mapping out the cosmic expansion history. Here, we discuss the application of the angular two-point correlation function, $w(\theta)$, to a sample of luminous red galaxies of the Sloan Digital Sky Survey (SDSS) and derive two new measurements of the BAO angular scale at $z = 0.235$ and $z = 0.365$. Since noise and systematics may hinder the identification of the BAO signature in the $w - \theta$ plane, we also introduce a potential new method to localize the acoustic bump in a model-independent way. We use these new measurements along with previous data to constrain cosmological parameters of dark energy models and to derive a new estimate of the acoustic scale $r_s$.}

%\abstract{}

\section{Introduction}
\label{sec:1}

Along with measurements of the luminosity of distant type Ia supernovae (SNe Ia) and the anisotropies of the cosmic microwave background (CMB), data of the large-scale distribution of galaxies have become one of the most important tools to probe the late-time evolution of the universe. This kind of measurement encodes not only information of the cosmic expansion history but also of the growth of structure, a fundamental aspect to probe different mechanisms of cosmic acceleration as well as to distinguish between competing gravity theories. In particular, recent measurements of a tiny excess of probability to find pairs of galaxies separated by a characteristic scale $r_s$ --  the comoving acoustic radius at the drag epoch -- was revealed in the two-point spatial correlation function (2PCF) of large galaxy catalogs. 

This BAO signature arise from competing effects of radiation pressure and gravity in the primordial plasma which is well described by the Einstein-Boltzmann equations in the linear regime~\cite{PeeblesY, SZ, BE, Hu-Dodelson, blake}. The first detections of the BAO scale in the galaxy distribution were obtained only in the past decade from galaxy clustering analysis of the Two Degree Field Galaxy Survey (2dFGRS)~\cite{cole05} and from the Luminous Red Galaxies (LRGs) data of the Sloan Digital Sky Survey (SDSS)~\cite{Eisenstein05}. More recently, higher-$z$ measurements at percent-level precision were also obtained using deeper and larger galaxy surveys \cite{Percival10, Beutler11, Blake11b, Sanchez12} (see \cite{eisensteinreview} for a recent review). 

The BAO signature defines a statistical standard ruler and provides independent estimates of the angular diameter distance $D_{\!A}(z)$ and the Hubble parameter $H(z)$ through the transversal ($dr_{\perp} = (1+z)D_A\theta_{BAO}$) and radial ($dr_{\parallel} = c\delta z/H(z)$) BAO modes, respectively. However, it is worth mentioning that the detection of the BAO signal through the 2PCF, i.e., using the 3D positions of galaxies, makes necessary the assumption of a fiducial cosmology in order  to transform the measured angular positions and redshifts  into  comoving distances. Such conversion  may  bias  the  parameter  constraints, as discussed in Refs. \cite{Eisenstein05, sanchez} (see also \cite{Salazar}).  

On the other hand, the calculation of the angular 2-point correlation function (2PACF), $w(\theta)$, involves only the angular separation $\theta$ between pairs, yielding information of $D_{A}(z)$ almost model-independently, provided that the comoving acoustic scale is known.  In order to extract information using 2PACF,  the galaxy sample is divided into redshift shells whose width has to be quite narrow ($\delta z \leq 10^{-2}$) to avoid large projection effects from the radial BAO signal. Another important issue in this kind of analysis is how to identify the actual BAO bump once the 2PACF is noisy and  usually exhibits more than one single bump due to systematic effects  present  in  the sample  (we refer the reader to \cite{gabriela} for a detailed discussion on this point). In what follows, we discuss the application of the 2PACF to large galaxy samples and introduce a potential new method to identify the BAO signature in a model-independent way. We exemplify the method with a sample of 105,831 LRGs from the seventh data release of the Sloan Digital Sky Survey (SDSS) and obtain two new measurements of $\theta_{BAO}$ at $z = 0.235$ and $z = 0.365$.

\section{The angular two-point correlation function}
\subsection{Theory}
\label{sec:2}

In the cosmological context, the {two-point correlation function}, $\xi(s)$, is defined as the excess probability of finding two pairs of galaxies at {a} given distance $s$. This function is  obtained by comparing the real catalog to random catalogs that follow the geometry of the survey~\cite{peebles,paddy_book}.  The most commonly used estimator of the 2PCF is the one proposed in Ref.~\cite{Landy}:
\begin{equation} \label{ls}
\xi(s) = \frac{DD(s) - 2 DR(s) + RR(s)}{RR(s)} \, , 
\end{equation}
where $DD(s)$ and $RR(s)$ correspond to the number of galaxy pairs with separation $s$ in real-real and random-random catalogs, respectively, whereas $DR(s)$ stands for the number of pairs with comoving separation $s$ calculated between a real-galaxy and a random-galaxy.
 %catalog. 

Assuming a flat universe, as indicated by recent CMB data~\cite{wmap9,planck}, the comoving distance $s$ between a pair of galaxies at redshifts $z_1$ and $z_2$ is given by 
\begin{equation}
s = \sqrt{r^2(z_1) \,+\, r^2(z_2) \,-\, 2 \,r(z_1) \,r(z_2)  \cos \theta_{12} \,\,} \, ,
\end{equation}
where $\theta_{12}$ is the angular distance between such pair of galaxies, and the radial distance between the observer and a galaxy at redshift $z_i$, $r(z_i) = c\int_{0}^{z_i}{dz/H(z, \bf{p})}$, depends on the  parameters ${\bf{p}}$  of the cosmological model adopted in the analysis. 

Similarly to the 2PCF, the 2PACF is defined as the excess joint probability that two point sources are found in two solid angle elements $d\Omega_1$ and $d\Omega_2$ with angular separation $\theta$ compared to a homogeneous Poisson distribution~\cite{peebles}. As mentioned earlier, this function can be used model-independently, considering only angular separations in narrow redshift shells of small $\delta z$ in order to avoid contributions from the BAO mode along the line of sight. The function $w(\theta)$ is calculated analogously to Eq. (\ref{ls}) with $s$ being replaced by $\theta$. The expected 2PACF, $w_{E}$, is given by~\cite{Carnero}
\begin{equation} \label{expected}
w_{E}(\theta, \bar{z}) = \int_0^\infty dz_1 \,\phi(z_1) \int_0^\infty dz_2 \,\phi(z_2) \,\xi_{E}(s, \bar{z}) \, , 
\end{equation}
where $\bar{z} \equiv (z_1 + z_2) / 2$, with $z_2 = z_1 + \delta z$, and  $\phi(z_i)$ is the normalised galaxy selection function at redshift $z_i$.  Note that, for narrow bin shells, $\delta z \approx 0$, so that $z_1 \approx z_2$ and $\xi_{E}(s, z_1) \simeq \xi_{E}(s, z_2)$. Therefore, one can safely consider that $\xi_{E}(s, \bar{z})$ depends only on the constant parameter $\bar{z}$, instead of on the variable $z$. 
The function $\xi_{E}(s, z)$ is given by~\cite{Sanchez11}
\begin{equation}\label{xi_e}
\xi_{E}(s,z)=\! \int_0^\infty \! \frac{dk}{2\pi^2} \, k^2 \, j_0(k s) \, b^2 \, P_m(k, z) \, , 
\label{eq:xiexp}
\end{equation}
where $j_0$ is the zeroth order Bessel function,  $P_m(k, z)$ is the matter power spectrum and $b$ is the bias factor. For shells of arbitrary $\delta z$, we refer the reader to~\cite{MSP}.

\subsection{Application to the data}

\begin{figure}[t]
\sidecaption[t]
\includegraphics[scale=.35]{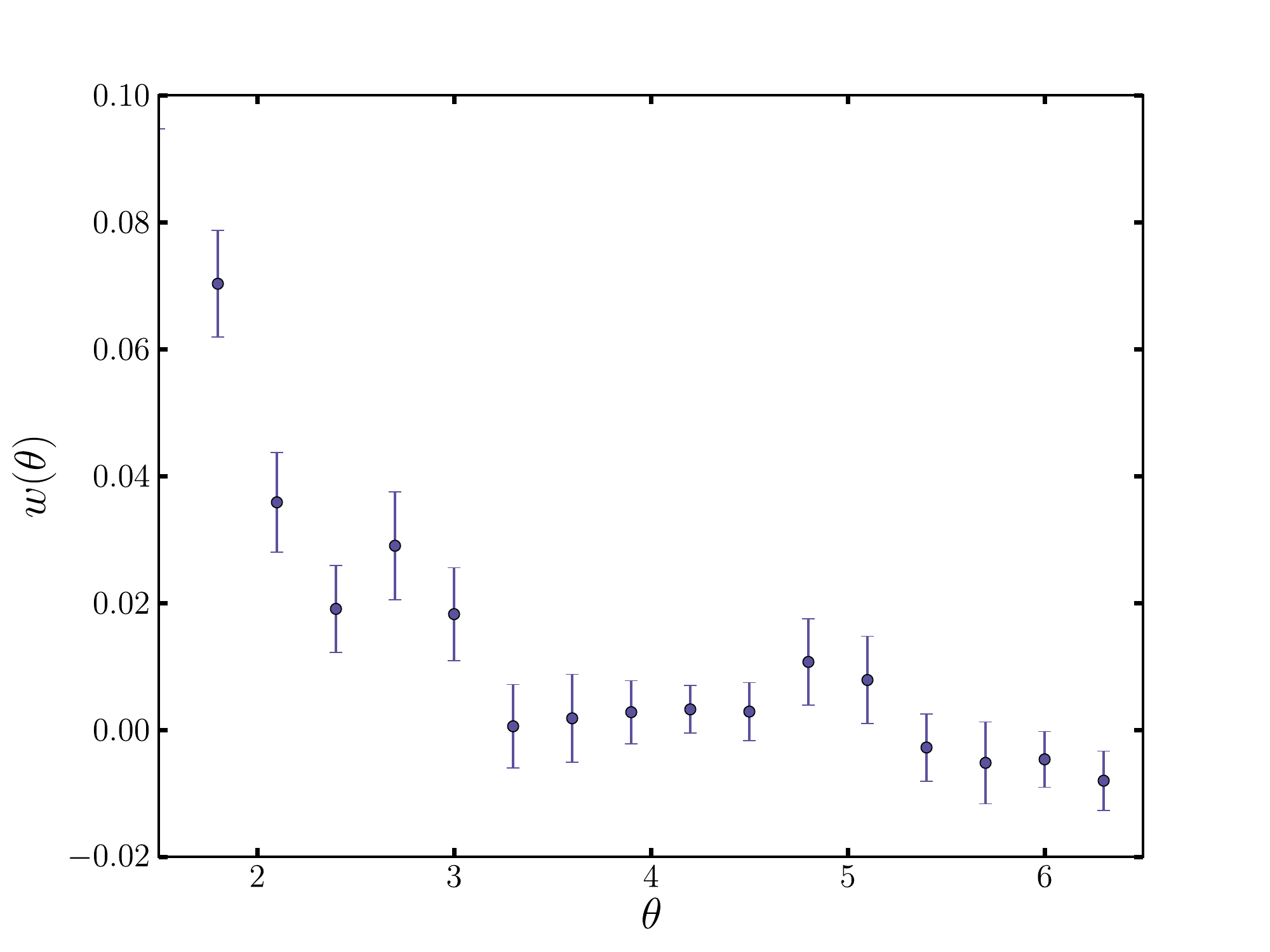}
\caption{An example of the 2PACF. As mentioned in the text, noise and systematics may give rise to bumps at different scales which hinder the identification of the BAO signature imprinted on the data. Clearly, two possible BAO bumps are observed at $\theta = 2.5^{\circ} - 3.0^{\circ}$ and $\theta = 4.8^{\circ} - 5.2^{\circ}$.}
\label{fig:1}       
\end{figure}

As illustrated in Fig. (\ref{fig:1}), the 2PACF derived from Eq. (\ref{ls}) usually exhibits more than {one} single {bump}  which, in general, is due to systematic effects present in the galaxy samples. In order to identify, among all bumps, which one corresponds to the real BAO scale, the usual procedure in the literature (see, e.g., \cite{Carnero}) is to compare the bump scales observed in the 2PACF with a cosmological model prediction obtained from Eqs. (\ref{expected}) and (\ref{xi_e}). Here, since we want to perform an analysis as model-independent as possible, we adopt the following criterium~\cite{gabriela}: if the BAO bump is present in the sample and is robust, then it will survive to changes in the galaxies angular coordinates by small and random amounts whereas the bumps produced by systematic effects will not.

We apply this procedure to a sample of 105,831 LRGs of the seventh public data release of the Sloan Digital Sky Survey distributed in the redshift interval $z = [0.16-0.47]$~\cite{dr7}. The transversal signatures as a function of redshift are obtained by dividing the data into two shells of redshift: $z = [0.20-0.27]$ and $z = [0.34-0.39]$, containing $19,764$ and $24,879$ LRGs, whose mean redshifts are $\bar{z} = 0.235$ and  $\bar{z} = 0.365$, respectively. The 2PACF as a function of the angular separation $\theta$ for this distribution of galaxies is shown in Figs. (\ref{fig:2}a) and (\ref{fig:2}b), where several bumps at different angular scales are observed.  However, using the criterium mentioned earlier, we calculate a number of 2PACF performing random displacements in the angular position of the galaxies, i.e., following Gaussian distributions with $\sigma = 0.25^{\circ}, 0.5^{\circ}$, and $   1.0^{\circ}$~\cite{gabriela}. Without assuming any fiducial cosmology, we find that only the bumps localised at $\theta_{\rm{FIT}} = 7.75^{\circ}$ and $\theta_{\rm{FIT}} = 5.73^{\circ}$ remain\footnote{Only for comparison,  the predictions of a flat $\Lambda$CDM cosmology, assuming $\Omega_m = 0.27$ and $ r_s=100 {\rm{Mpc/h}}$, are $\theta_{\rm{BAO}}(0.235) = 8.56^{\circ}$ and $\theta_{\rm{BAO}}(0.365) = 5.68^{\circ}$.}. 

Another potential method for identifying the real BAO scale may be achieved by counting the number of neighbours of the sub-sample of galaxies contained in each shell. To make this explicit, we first select all the pairs contributing to the BAO bump at  $\theta_{\rm{FIT}}{(\bar{z} = 0.235)} = 7.75^{\circ}$ in the 2PACF, i.e., pairs of galaxies that have an angular separation distance in the interval  $[7.6^{\circ},7.8^{\circ}]$. Within this set, we analyse the repetition rate: given a galaxy, we compute how many galaxies (that is, neighbours) are apart by an angular distance between $7.6^{\circ}$ to $7.8^{\circ}$ from that galaxy. The same procedure is applied to the galaxies in the interval $[5.6^{\circ},5.8^{\circ}]$, where a bump is also observed at $\simeq 5.7^{\circ}$ (Fig. {\ref{fig:2}a). The result is shown in Fig. {\ref{fig:3}}, where we find that, while the mean number of neighbours in the latter interval is $N_n = 16 \pm 7$, the value of $N_n$ in the former (the BAO bump) is $20 \pm 9$ (Fig. (\ref{fig:3}a)). By similar proceeding, we also find this same characteristic in the sub-sample of galaxies contained in the shell $z = [0.34-0.39]$, as shown in Fig. (\ref{fig:3}b). It is worth mentioning that similar results are also found for the samples of LRGs of the tenth and eleventh data release of the SDSS. In other words, it seems that galaxies around the BAO bumps have a larger number of neighbours than those around non-BAO bumps. If confirmed in other galaxy samples, this property could also be used to identify the real BAO scales in a model-independent way.

\begin{figure*}[t]
\includegraphics[scale=.30]{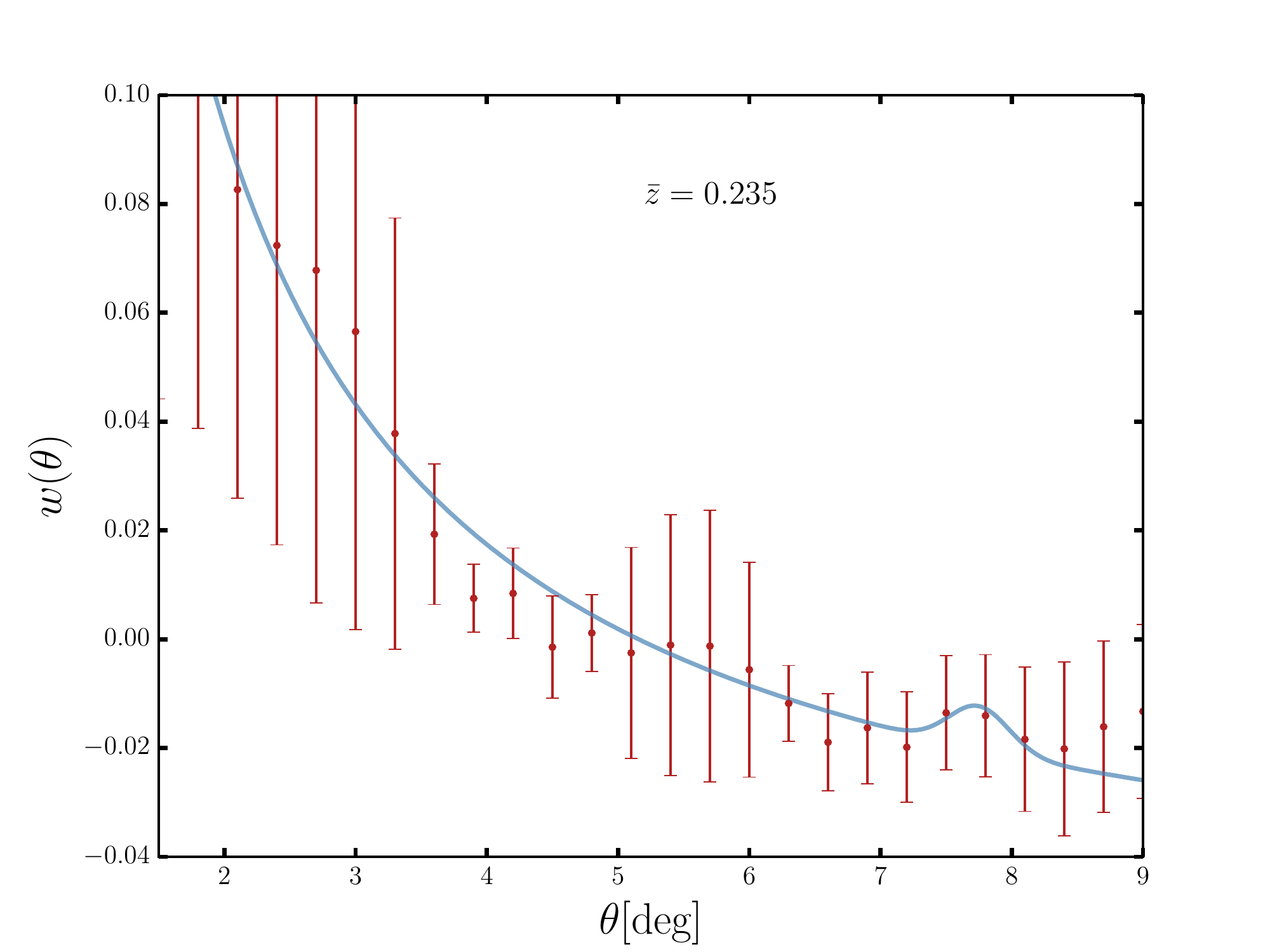}
\includegraphics[scale=.30]{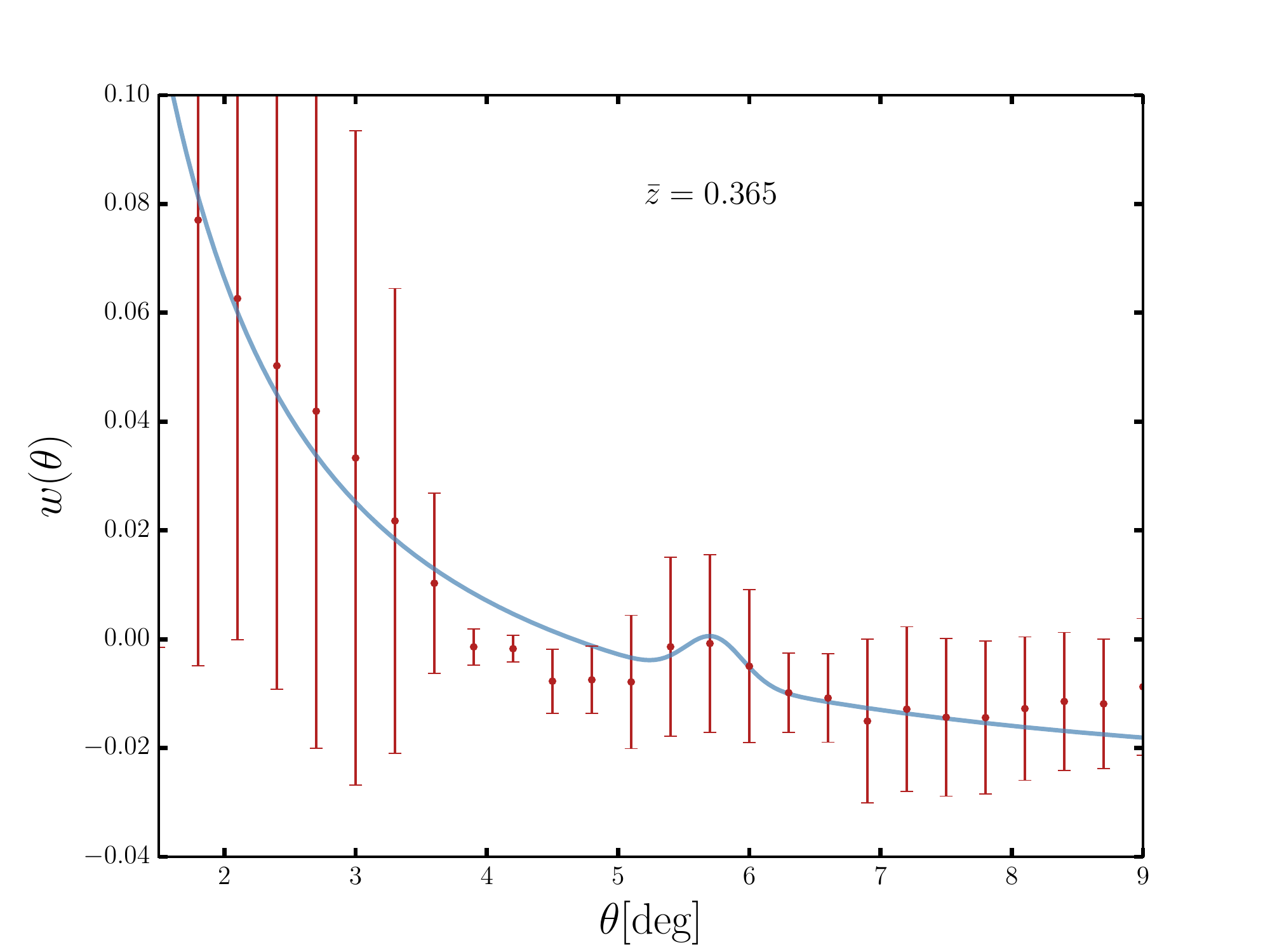}
\caption{The 2PACF for two redshift intervals obtained from the LRGs sample of the SDSS-DR7 (red bullets).  The continuous line is derived from the 2PACF parameterisation proposed in Ref.~\cite{Sanchez11}.}
\label{fig:2}       
\end{figure*}

\begin{figure*}[t]
\includegraphics[scale=.305]{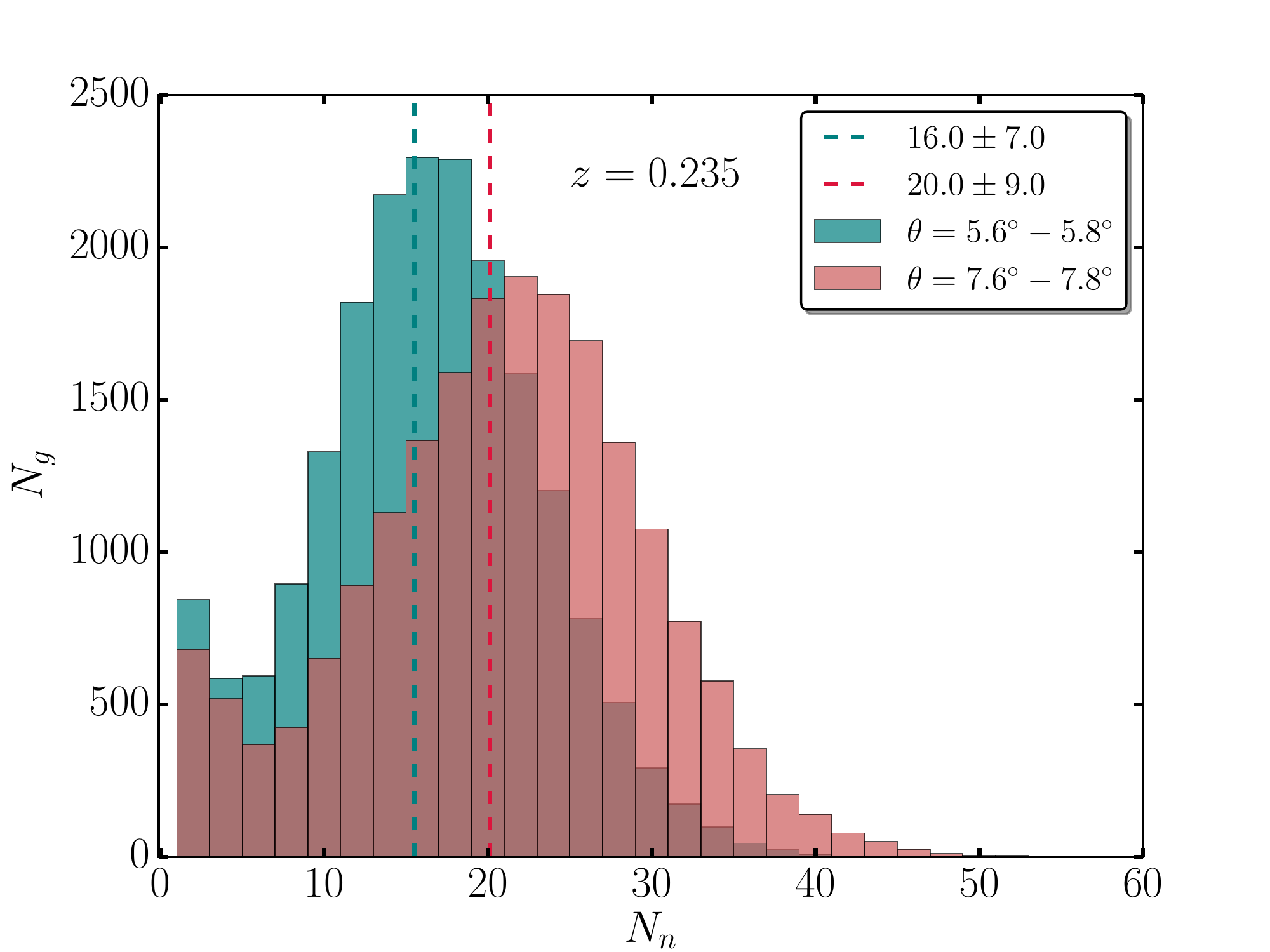}
\includegraphics[scale=.305]{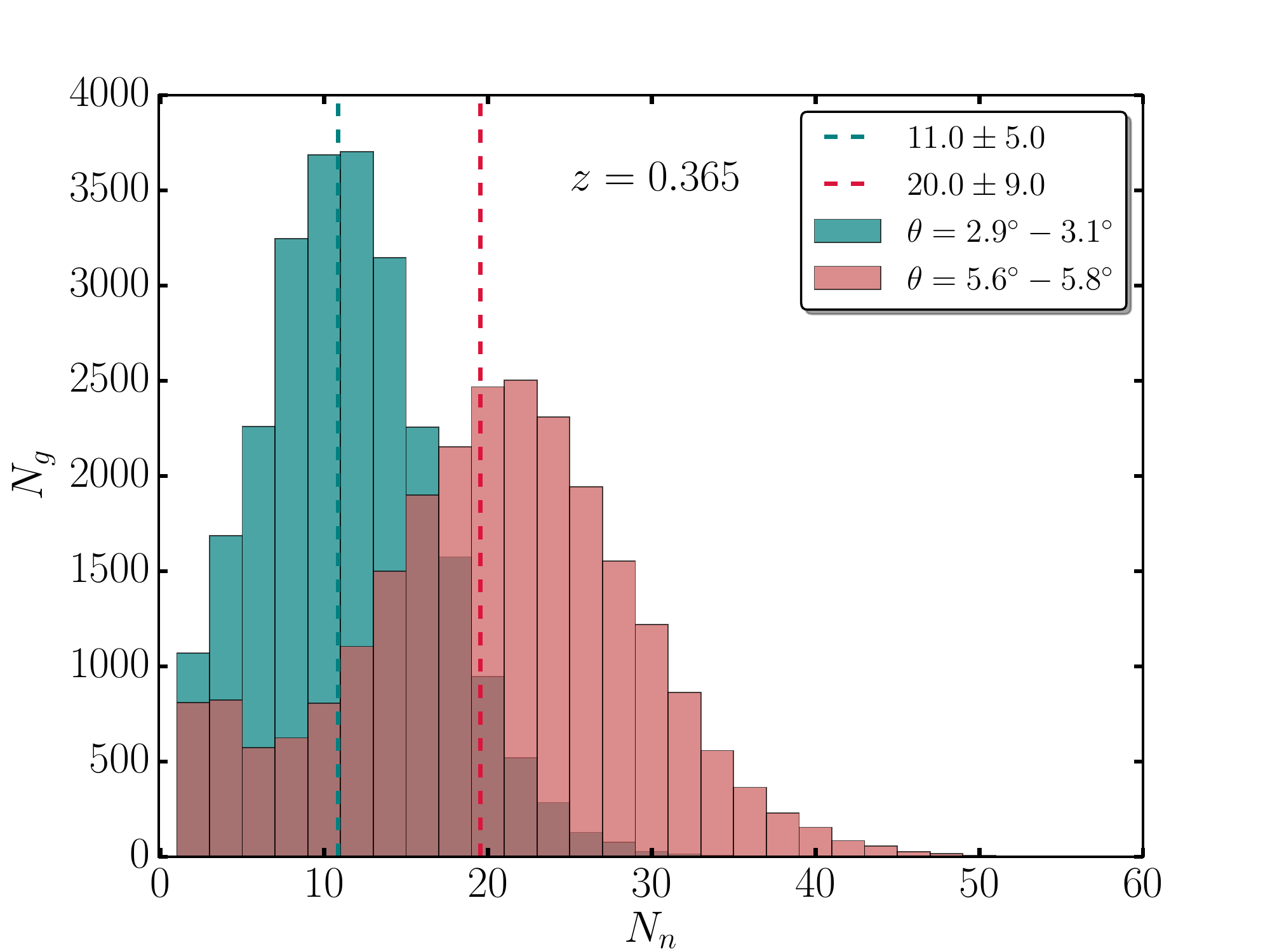}
\caption{{\it{Left:}} Histograms of the number of neighbours for galaxy pairs in the intervals $[7.6^{\circ}, 7.8^{\circ}]$ and $[5.6^{\circ}, 5.8^{\circ}]$ $(\bar{z} = 0.235)$. The horizontal axis, $N_n$, represents the number of neighbours of each galaxy whereas vertical dashed lines indicate the mean value of each distribution. {\it{Right:}} The same as in the previous panel for  galaxy pairs in the intervals  $[5.6^{\circ}, 5.8^{\circ}]$ and $[2.9^{\circ}, 3.1^{\circ}]$ $(\bar{z} = 0.365)$.}
\label{fig:3}       % Give a unique label
\end{figure*}

After {finding} the real BAO signature, we obtain the angular BAO scale using the method of Ref.~\cite{Sanchez11}, which {parameterises} the 2PACF as a sum of a power law, describing the continuum, and a Gaussian peak, which describes the BAO bump, i.e.,
\begin{equation} 
w_{FIT}(\theta) = A  + B \theta^\nu + C e^{-\frac{(\theta-\theta_{FIT})^2}{2 \sigma_{FIT}^2}} \, ,
\label{eq:fit}
\end{equation}
where $A,B,C,\nu$, and $\sigma_{FIT}$ are free parameters, $\theta_{FIT}$ defines the position of the acoustic scale and $\sigma_{FIT}$ gives a measure of 
the width of the bump. Note that, if $\delta z = 0$, the true BAO scale $\theta_{BAO}$ and $\theta_{FIT}$ would coincide. However, for $\delta z \neq 0$  projection effects due to the width of the redshift {shells} must be taken into account (see Fig. (\ref{fig:4}a)). Therefore, assuming a fiducial cosmology, the function $w_{E}(\theta, \bar{z})$, given by Eqs.~(\ref{expected}) and (\ref{eq:xiexp}), has to be calculated for both $\delta z = 0$ and $\delta z \neq 0$ in order to compare the position of the peak in the two cases. This allows to find a correction factor $\alpha$ that, given the value of $\theta_{FIT}$ estimated from Eq. (\ref{eq:fit}), provides the true value for $\theta_{BAO}$. To perform the calculation of $\alpha$, we assume the standard $\Lambda$CDM cosmology\footnote{For narrow redshift shells, such as the ones considered in this analysis ($\delta z \sim 10^{-2}$), it can be shown that the correction factor depends weakly on the cosmological model adopted (see Fig. 3 of \cite{Sanchez11}).} and also consider the correction due to the photometric error of the sample by following Ref.~\cite{Carnero}. After all these corrections, the two new measurements of the BAO angular scales at $\theta_{\rm{BAO}} = [9.06 \pm 0.23]^{\circ}$ ($\bar{z} = 0.235$) and $\theta_{\rm{BAO}} = [6.33 \pm 0.22]^{\circ}$ ($\bar{z} = 0.365$) are obtained.

\begin{figure*}[t]
\includegraphics[width=4.6cm,height=5.1cm]{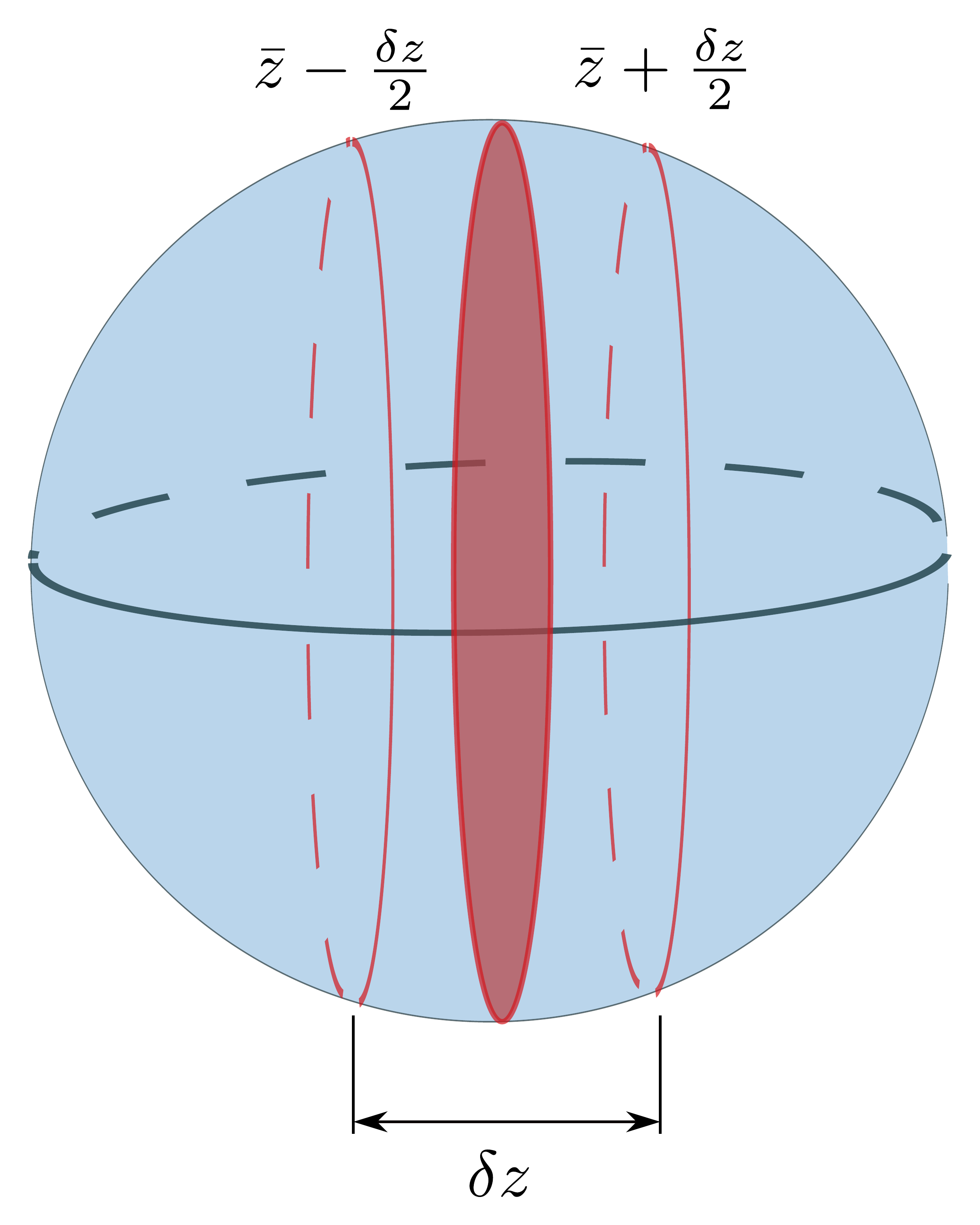}
\hspace{1.0cm}
\includegraphics[scale=.32]{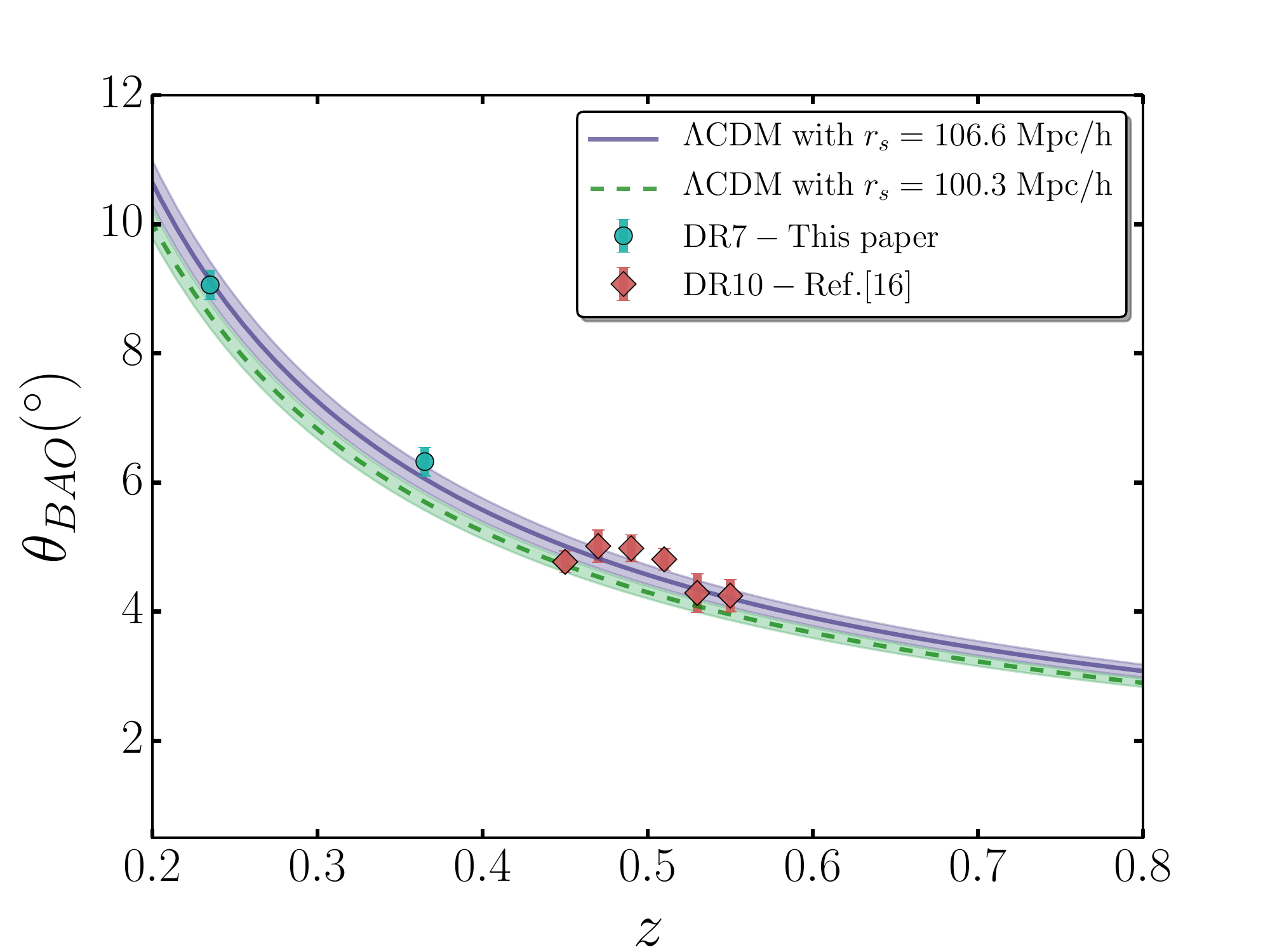}
\caption{{\it{Left:}} Projection effect: Let us consider that the sphere represents the BAO signature. When the rings at $\bar{z} \pm \delta z/2$ are projected into the the shaded region ($\bar{z}$), the signature appears smaller than true one. Therefore, this effect produces a shift of the BAO peak to  lower values. {\it{Right:}} The angular BAO scale as a function of redshift. The green data points correspond to the two measurements obtained in this analysis whereas the red ones are the data points obtained in Ref.~\cite{gabriela} using the LRGs of the SDSS-DR10. The curves stand for the $\Lambda$CDM prediction with the acoustic scale fixed at the WMAP9 and Planck values.}
\label{fig:4}       % Give a unique label
\end{figure*}

\section{Cosmological constraints}

We present cosmological parameter fits to the BAO data derived in the previous section along with the data set obtained in Ref.~\cite{gabriela} (see Fig.~\ref{fig:4}). The angular scale $\theta_{BAO}$ is related to the acoustic radius $r_s$ and the angular diameter distance $D_A(z)$ {through}
\begin{equation} \label{thetaz}
 \theta_{BAO} (z)  = \frac{r_s }{(1+ z) D_A(z)}\,,
\end{equation}
where 
\begin{equation}
D_A(z) = {c \over (1+z)}\int_{0}^{z}{dz' \over H(z,\mathbf{p})}\;.
\end{equation}
In what follow, we explore three classes of cosmological models, starting from the minimal $\Lambda$CDM cosmology. We also consider a varying dark energy model whose equation-of-state parameter evolves as $w = w_0 + w_a[1-a/(2a^2 - 2a + 1)]$~\cite{BA} and a particular case when $w_0 < 0$ and $w_a = 0$ ($w$CDM) (for a discussion on theoretical models of dark energy, see~\cite{paddy}). In our analysis, we use the WMAP9 final estimate of the comoving acoustic radius at the drag epoch, $r_s = 106.61 \pm 3.47$$h^{-1}$ Mpc~\cite{wmap9}. Plots of the resulting cosmological constraints are  shown in Fig.~(\ref{fig:5}). Although  the $\theta_{\rm{BAO}}$ data  alone  (gray  contours)  are  consistent with a wide interval of $w_0$ and $w_a$ values, their combination with the CMB data limits considerably the range of $w$, favouring values close to the cosmological constant limit $w = -1.0$. This can be seen when we combine the BAO data points with CMB measurements of the shift parameter (red contours), defined as ${\cal{R}} = \sqrt{\Omega_m}\int^{z_{ls}}_{0}H_0/H(z) dz$, where $z_{ls}$ is the redshift of the last scattering surface. In order to avoid double counting of information with the $r_s$ value from WMAP9 used in the BAO analysis, we use ${\cal{R}} = 1.7407 \pm 0.0094$, as given by the latest results from the Planck Collaboration~\cite{planck}. The joint results (brown contours) improve significantly the cosmological constraints, providing the values shown in Table I.

\begin{figure*}[t]
\includegraphics[scale=.286]{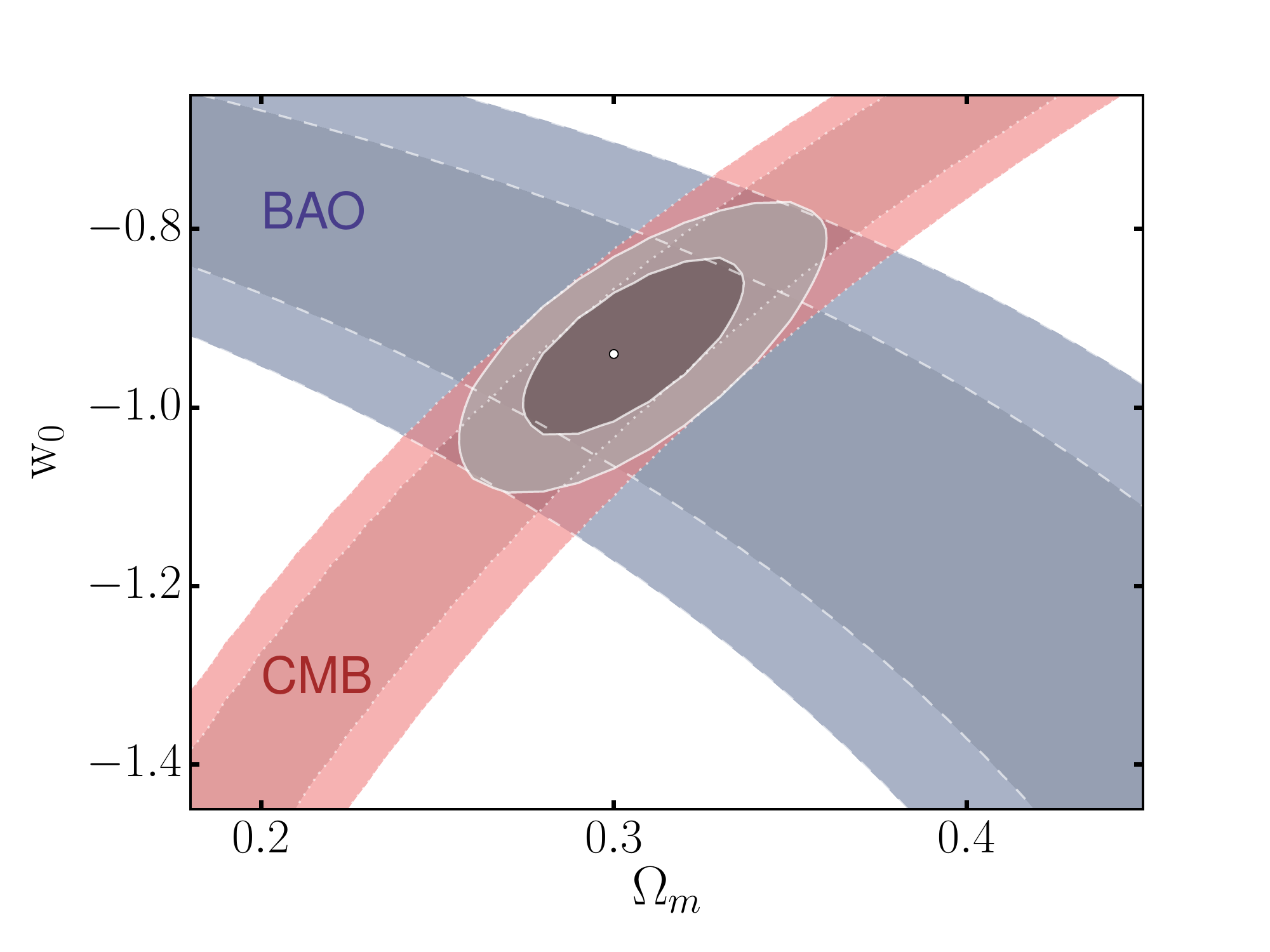}
\includegraphics[scale=.286]{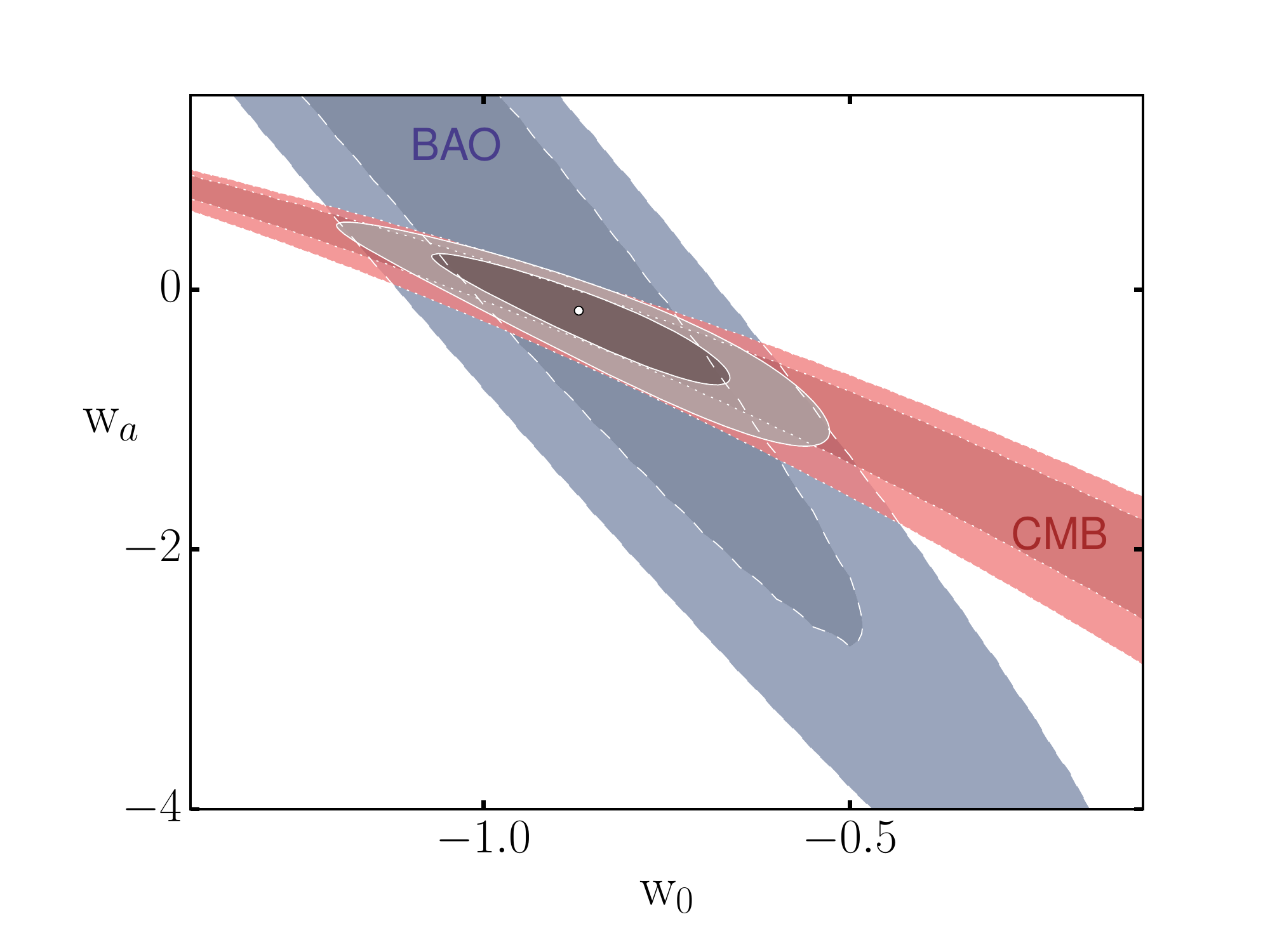}
\caption{{\it{Left:}} Confidence regions in the $\Omega_m - w_0$ plane. The grey contours correspond to the region allowed by the current $\theta_{\rm{BAO}}$ data (SDSS - DR7/DR10) whereas the red contours are obtained from CMB data. The combination between $\theta_{\rm{BAO}}$  and CMB limits considerably the allowed interval of the cosmological parameters. {\it{Right:}} The same as in the previous panel for the $w_0 - w_a$ plane.}
\label{fig:5}       % Give a unique label
\end{figure*}

\begin{table}[b]
\caption{Constraints on model parameters. The error bars correspond to 1$\sigma$.}
\label{tab:1}    
\begin{tabular}{p{2cm}p{2.4cm}p{2cm}p{4.9cm}}
\hline\noalign{\smallskip}
Model & $\Omega_m$ & $w_0$ & $w_a$  \\
\noalign{\smallskip}\svhline\noalign{\smallskip}
$\Lambda$CDM & $0.27 \pm 0.04$  & -1 & - \\
$w$CDM & $0.30\pm 0.02$ & $-0.94 \pm 0.06$ & - \\
$w(z)$CDM & $0.29 \pm 0.03$ & $-0.87 \pm 0.13$ & $-0.16 \pm 0.32$\\
\noalign{\smallskip}\hline\noalign{\smallskip}
\end{tabular}
%$^a$ Marginalized over the matter density parameter.
\end{table}

Finally, it is important to observe that,  given a cosmological model, estimates of the acoustic scale $r_s$ can be obtained directly from Eq. (\ref{thetaz}), i.e., independently of CMB data. Using the data set shown in Fig. (\ref{fig:4}) and assuming the $\Lambda$CDM scenario, we find $r_s = 103.6 \pm 4.1$ $h^{-1} \rm{Mpc}$, which is in good agreement with both the WMAP9 and Planck estimates as well as with the value obtained in~\cite{Heavens}.

\section{Conclusions}

Sourced by the initial density fluctuations, the primordial photon-baryon plasma supports the propagation of acoustic  waves until the decoupling of photons and baryons. These oscillations imprinted a preferred clustering scale in the large scale structure of the universe which have been detected either as a peak in the real space correlation function or as a series of
peaks in the power spectrum. In this paper, we have discussed the application of the angular two-point correlation function to a sample of 105,831 LRGs of the seventh public data release of the Sloan Digital Sky Survey (SDSS) distributed in the redshift interval $z = [0.16-0.47]$~\cite{dr7}. Differently from analysis that use the spatial correlation function, $\xi(s)$, where the assumption of a fiducial cosmology is necessary in order  to transform the measured angular positions and redshifts into comoving distances, the calculation of the 2PACF, $w(\theta)$, involves only the angular separation $\theta$ between pairs, yielding measurements of the BAO signal almost model-independently.

After identifying the BAO peaks using the method of Ref.~\cite{gabriela} and introducing a new potential method based on the mean number of neighbours of the galaxies contained in the redshift shells, we have derived two new measurements of the BAO angular scale: $\theta_{\rm{BAO}} = [9.06 \pm 0.23]^{\circ}$ ($\bar{z} = 0.235$) and $\theta_{\rm{BAO}} = [6.33 \pm 0.22]^{\circ}$ ($\bar{z} = 0.365$). As shown in Fig.~(\ref{fig:4}b), these low-$z$ measurements are important to fix the initial scale in the $\theta_{\rm{BAO}} - z$ plane, which improves the constraints on cosmological parameters. Along with six measurements of $\theta_{\rm{BAO}}(z)$ recently obtained in Ref.~\cite{gabriela}, we have used the data points derived in this analysis to constrain different dark energy models. We have found a good agreement of these measurements with the predictions of the standard $\Lambda$CDM model as well as with some of its simplest extensions. Assuming the standard cosmology, we have also derived a new estimate of the acoustic scale, $r_s = 103.6 \pm 4.1$ $h^{-1} \rm{Mpc}$ ($1\sigma$). This value is obtained from the distribution of galaxies only and is in good agreement with recent estimates from CMB data assuming the $\Lambda$CDM cosmology.

\begin{acknowledgement}
Jailson S. Alcaniz dedicates this contribution to Prof. T. Padmanabhan with affection and profound admiration. May he continue to inspire us with new ideas for many more years. Happy 60th birthday, Paddy! 

The authors acknowledge support from the Conselho Nacional de Desenvolvimento Cient\'{\i}fico e Tecnol\'ogico (CNPq) e Funda\c{c}\~ao Carlos Chagas de Amparo \`a Pesquisa do Estado do Rio de Janeiro (FAPERJ). Joel C. Carvalho is also supported by the DTI-PCI/Observat\'orio Nacional program of the Brazilian Ministry of Science, Technology and Innovation (MCTI). 
\end{acknowledgement}

\end{document}